\documentclass[aps,twocolumn,amsfonts,showpacs,pra,amsmath,amssymb,superscriptaddress]{revtex4-1}
\usepackage{graphicx, color}
\usepackage{hyperref}
\hypersetup{colorlinks=true,linkcolor=blue,citecolor=blue,urlcolor=blue}
\usepackage{dcolumn}
\usepackage{bm}
\usepackage{mathtools}
\usepackage{newtxtext}
\usepackage{newtxmath}
\usepackage[version=3]{mhchem}
\usepackage[normalem]{ulem}
\usepackage{lipsum}
\usepackage{braket}
\usepackage{comment}

\newcommand{\red}{\textcolor{black}}

\newcommand{\trl}{{}^{t}{\!\!\,}}

\begin{document}

\title{Photoinduced pseudospin-wave emission from \\
charge-density-wave domain wall with superconductivity}

\newcommand{\TohokuUniv}{Department of Applied Physics, Tohoku University, Sendai 980-8579, Japan}

\author{Yukihiro Matsubayashi}
\email{yukihiro.matsubayashi.s8@dc.tohoku.ac.jp}
\affiliation{\TohokuUniv}

\author{Yusuke Masaki}
\email{yusuke.masaki.c1@tohoku.ac.jp}
\affiliation{\TohokuUniv}
\affiliation{Research and Education Center for Natural Sciences, Keio University, Hiyoshi 4-1-1, Yokohama, Kanagawa 223-8521, Japan}
\author{Hiroaki Matsueda}
\affiliation{\TohokuUniv}
\affiliation{Center for Science and Innovation in Spintronics, Tohoku University, 2-1-1 Katahira, Aoba, Sendai, Miyagi 980-8577 Japan}
\date{\today}

\begin{abstract}
We study photoinduced dynamics triggered by an inhomogeneity due to competition between charge density waves (CDWs) and superconductivity. 
As a simple example, we consider the superconducting (SC) interface between two CDW domains with opposite signs. 
The real-time dynamics are calculated within the time-dependent Hartree--Fock--Bogoliubov framework,
where the order parameter dynamics and the nonequilibrium quasiparticle distribution functions are studied.
We also calculate the various dynamical response functions within a generalized random phase approximation. 
Through comparisons between the real time dynamics and the analysis of the response functions, it is found that the photo-driven SC interface can emit collective modes of the SC order parameter. This is analogous to the spin wave emission from the magnetic domain wall in an antiferromagnet, particularly in the case of a low driving frequency, where the order parameters can be mapped onto the pseudospin picture. 
In the high-frequency case, we find a domain wall melting caused by changes in the quasiparticle distribution, which induces superconductivity in the whole system.  
\end{abstract}

\maketitle

\section{INTRODUCTION}
Recent advances in terahertz laser technologies have enabled us to access the low-energy scales important for elucidating the fundamental properties of solids, e.g., the energy scales of phonons, excitons, plasmons, magnons, superconducting (SC) gaps and density waves~\cite{Giannetti2016_Rev_NoneqCuprate, Ishihara2019_Rev, DeLaTorre2021_Rev}.
Terahertz lasers can be used to investigate not only the linear response of matter but also the nonlinear response and highly nonequilibrium phenomena thanks to its strong intensity.
For example, observation of the Higgs mode, which is the amplitude mode of the SC pair potential, has been achieved by non-adiabatic excitation of the Bardeen--Cooper--Schrieffer (BCS) ground state with an ultrafast intense terahertz laser~\cite{Matsunaga2013,Matsunaga2014,Katsumi2018,Shimano2020_Rev}.
Among the various research topics, optical control of superconductivity has crucial importance for both fundamental physics and technological applications.
Intensive studies have been conducted to enhance the transition temperature $T_c$ and the SC order \cite{Kaiser2014, Nicoletti2014, Casandruc2015, Sentef2015, Knap2016, Murakami2017, Ido2018_dSC_CDW, Kennes2019, Wang2021}
and to realize SC states different from the equilibrium one, such as photoinduced topological superconductivity~\cite{Takasan2017, Chono2020, Yanase2022, Wenk2022}.

Electronic phases competing with superconductivity present the possibility of indirect photo-control of the SC phase. 
Such orders can be inferred to exist in various materials such as transition metal dichalcogenides~\cite{Klemm2015, Petkov2020, Pasztor2021, Xu2021_Rev, Ren2022} and cuprate supercondoctors \cite{Demler2004, Wise2008, Subir2012, Eduardo2014, Croft2014, Fradkin2015_Rev, Campi2015, Lu2015, Peng2016, Jang2017, Uchida2021, Imada2021_SCCDW_Rev}.
Numerous experiments and theoretical studies have also indicated the importance of competing multiple orders with inhomogeneity~\cite{Emery1993, Sboychakov2007, Sipos2008, Banerjee2018, Seo2018, Wen2019, Kagan2021, Chen2022_EHM}.
In addition, it is considered that inhomogeneity acts as a trigger in the initial process of the photoinduced phase transition \cite{Tokura2006, Seo2018, Zong2019}.
Further investigations into the nonequilibrium dynamics of such systems are required in order to pave a way for discovering nontrivial phenomena.

The attractive Hubbard model has been widely used to study the nonequilibrium dynamics of superconductors~\cite{Scalettar1989, Sentef2017_AttHubb, Chern2019_EoM, Vitali2020, Fontenele2022}.
This model has received much attention not only as a simple toy model of superconductivity but also in terms of its realization with ultracold atom% due to developing experimental techniques 
~\cite{Bloch2008_Rev, Schafer2020_Rev, Mitra2018, Chan2020, Gall2020}.
In this model, the superconductivity and charge density wave (CDW) are degenerate at half-filling~\cite{Dichtel1971,Nagaoka1974,Mertsching1977,Robaszkiewicz1981a,Robaszkiewicz1981,Micnas1990,Yang1990}. 
Laser control of superconductivity has been proposed for both the attractive and repulsive Hubbard model by suppressing the CDW order and charge inhomogeneity in recent theoretical studies~\cite{Sentef2017_AttHubb, Fujiuchi2020, Ido2018_dSC_CDW}.
These studies highlight the importance of degenerate orders when using laser irradiation to control electronic phases.

In this paper, we investigate the photoinduced nonequilibrium dynamics of a non-uniform system composed of SC and CDW orders.
To describe such a system, we use the extended attractive Hubbard model (EAHM) with an onsite attractive interaction $U<0$ and nearest-neighbor repulsive interaction $V>0$ on a square lattice.
The EAHM is known to host a variety of exotic SC phases such as $s$-wave, $p$-wave and $d$-wave symmetries \cite{Nayak2018, Leridon2020, Hutchinson2020, Chen2022_EHM}.
At half-filling, because $V > 0$, the CDW order is stabilized as the ground state, and the SC orders are destabilized.
In order to drive the dynamics of superconductivity, we consider a non-uniform system with an $s$-wave SC interface sandwiched by two CDW domains with opposite signs.
This is a simple setup where both superconductivity and CDW exist.
The time evolution in the laser field is calculated within the mean-field approximation.
We classified our results by the driving frequency of the laser $\omega_{\textrm{ext}}$ and the CDW gap $\omega_{\varg}$: (i) $\omega_{\textrm{ext}} \ll \omega_{\varg}$ and (ii) $\omega_{\textrm{ext}} \sim \omega_{\varg}$.
In case~(i), we find a collective mode emission from the domain wall.
The collective mode can be interpreted as a pseudospin wave, where the pseudospin represents the CDW and SC order parameters. 
In case~(ii), the CDW domains and the SC domain wall melt over time through the quasiparticle excitation.
On the other hand, the emission of the SC collective mode from the domain wall induces superconductivity throughout the system.
Their dynamics are discussed by analyzing the time-dependent quasiparticle population and pair potential.

This paper is organized as follows: Sec.~II introduces the EAHM and the time-dependent calculation method within the Hartree--Fock--Bogoliubov approximation.
It also introduces the charge and pair correlation functions.
The correlation functions are derived by linear response theory and calculated in the random phase approximation (RPA).
Section~III~A explains the non-uniform self-consistent solution and its description in terms of pseudospins.
Section~III~B discusses the collective mode emission from the domain wall under an electric field oscillating at a low frequency.
In Sec.~III~C, we discuss the nonequilibrium dynamics induced by the resonant excitation.
Finally, Sec.~IV presents a summary and discussion of the overall results of the paper. Note that this paper uses the unit $\hbar = |e| = c = k_{\mathrm{B}} = 1$.

\section{MODEL AND METHODS}
In this section, we introduce the model Hamiltonian and the mean field approximation with its self-consistent condition. Then, we derive a set of equations of motion for the time evolution driven by an applied external field. In subsect.~C, we develop a self-consistent linear response theory, which tells us information about the collective mode of the system from the imaginary part of the dynamical susceptibility. 

\subsection{Extended attractive Hubbard model}
As a minimal model of the system with the superconductivity and CDW, we introduce the extended attractive Hubbard model on a square lattice: 
\begin{align}
    H &=
    \sum_{i, j, \sigma} \mathcal{J}_{ij} c_{i\sigma}^\dagger c_{j\sigma} + U\sum_i n_{i\uparrow} n_{i\downarrow} \nonumber \\
    & \hspace{5em} + \dfrac{1}{2} \sum_{i, j} V_{ij} n_{i} n_{j}
    - \mu\sum_i n_i
    , \label{eq:ham1}
%    H &=
%    \sum_{i, j, \sigma} \mathcal{J}_{ij} c_{i\sigma}^\dagger c_{j\sigma} + \sum_i \left(U n_{i\uparrow} %n_{i\downarrow} - \mu n_i\right) \nonumber \\
%     & \hspace{2em} + \dfrac{1}{2} \sum_{i, j} V_{ij} n_{i} n_{j}~~~
%     \Bigl(n_i = \sum_{\sigma}n_{i\sigma} = \sum_{\sigma}c_{i\sigma}^{\dagger}c_{i\sigma}\Bigr)
%    , \label{eq:ham1}
%    H =
%    \sum_{i, j}\left(\sum_{\sigma} \mathcal{J}_{ij} c_{i\sigma}^\dagger c_{j\sigma} + \dfrac{V_{ij}}{2} n_{i} n_{j}\right)
%    + \sum_i \left(U n_{i\uparrow} n_{i\downarrow}
%    - \mu n_i\right)
%    , \label{eq:ham1}
\end{align}
where $c_{i\sigma}^\dagger\,(c_{i\sigma})$ is the creation (annihilation) operator at site $i$ with spin $\sigma$, %, and $\braket{i,j}$ denotes summation over the system, 
and $n_i = \sum_{\sigma}n_{i\sigma}=\sum_{\sigma}c_{i\sigma}^\dagger c_{i\sigma}$ is the number operator at site $i$.
Here, $i$ identifies the two-dimensional lattice site $\bm{r}_i = (i_x, i_y)$ and so does $j$.
We focus on the half-filling case by adjusting the chemical potential $\mu$.
The first term is the hopping Hamiltonian, where the hopping parameter without any external field is given by $\mathcal{J}_{ij} = J (< 0)$ for nearest-neighbor sites $i$ and $j$ and otherwise $0$, i.e., $\mathcal{J}_{ij} = J \delta_{|\bm{r}_i -\bm{r}_j|, 1}$. 
%accounts for nearest-neighbor hopping, while t
The second term describes the on-site attractive interaction for $U < 0$. 
%The second term with $U<0$ accounts for the on-site attractive interaction. % for $U < 0$. 
The third term describes the nearest-neighbor interaction for $V_{ij}=V \delta_{|\bm{r}_i -\bm{r}_j|, 1}$, which lifts the degeneracy between the SC and CDW states~\cite{Micnas1990}. When the nearest-neighbor interaction $V > 0$ ($V < 0$), the CDW (SC) phase has lower energy. 
%For a homogeneous system without any external field, the hopping parameter can be taken as $J_{ij} = J ( < 0)$. 
For the total number of the unit cells $N$, the Fourier transform $c_{i\sigma} = 1/\sqrt{N} \sum_{\bm{k}} e^{i\bm{k}\cdot\bm{r}_i} c_{\bm{k}\sigma}$ leads to the following representation of the Hamiltonian:  
\begin{equation}
    \begin{split}
        H = 
        \sum_{\bm{k} \sigma} \epsilon_{\bm{k} \sigma} n_{\bm{k}\sigma}
        + \dfrac{1}{2N} \sum_{\bm{k} \sigma\sigma'}
        \delta_{\sigma,\bar{\sigma}'} (U + V_{\bm{k}}) \rho_{-\bm{k}\sigma} \rho_{\bm{k}\sigma'},
    \end{split}
\end{equation}
where
$n_{\bm{k}\sigma}$ $=$ $c_{\bm{k}\sigma}^\dagger c_{\bm{k}\sigma}$, 
$\rho_{\bm{q}\sigma}$ $=$ $\sum_{\bm{k}} c_{\bm{k}\sigma}^\dagger c_{\bm{k}+\bm{q}\sigma}$, 
$\epsilon_{\bm{k}}$ $=$ $2J[ \cos(k_x)$ $+$ $\cos(k_y) ]$, $V_{\bm{k}}$ $=$ $2V[ \cos(k_x) + \cos(k_y) ]$.
The mean-field approximation of the Hamiltonian~\eqref{eq:ham1} is performed as follows:
The %second 
$U$ term takes into account the Hartree--Fock terms as well as the anomalous average, that is, the so-called Bogoliubov term. The $V$ term takes into account only the Hartree term in order to exclude the bond order wave and the $d$-wave superconductivity for simplicity. By using the mean fields 
$\braket{n_{i\sigma}}$ and
$\Delta_i \equiv \braket{c_{i\downarrow} c_{i\uparrow}}$, the %second and third terms 
interaction terms reduce to 
\begin{equation}
\begin{split}
    \sum_i (U\Braket{n_{i\bar{\sigma}}} - \mu) n_{i\sigma}
    + U \sum_i (\Delta_i c_{i\uparrow}^\dagger c_{i\downarrow}^\dagger + \mathrm{H.c.})
    + V \sum_{\braket{i, j}} \Braket{n_i} n_j .
\end{split}
\end{equation}
Hence, the EAHM can be rewritten in a quadratic form:
\begin{equation}
    H
    \simeq
    \vec{C}^\dagger H_{\mathrm{BdG}} \vec{C},    
\end{equation}
where $\vec{C}^{\dagger}=(c_{1\uparrow}^\dagger, \dots ,c_{N\uparrow}^\dagger, c_{1\downarrow},\dots, c_{N\downarrow} )$.
The static structures of the mean fields, $\braket{n_i}$ and $\Delta_i$, and the electron states are determined self consistently. From the mean-field Hamiltonian with a set of the mean fields, we obtain the one particle eigenenergies and eigenstates, which determine a new set of mean fields. We repeat this iterative procedure until the largest error in the updates becomes less than $\varepsilon_{\mathrm{err}} = 10^{-8}$.

\subsection{Equation of motion}
In order to calculate the real-space dynamics, we introduce normal and anomalous density matrices $\mathscr{G}_{i\sigma;j\sigma'}=\braket{c^\dagger_{j\sigma'} c_{i\sigma}}$, $\mathscr{F}_{i\sigma;j\sigma'}=\braket{c_{j\sigma'} c_{i\sigma}}$.
The time evolution is calculated on the basis of the equation of motion for the density matrices given by
\begin{equation}
\begin{split}
\label{eq:dGdt}
    -i\dfrac{d}{dt} \mathscr{G}
    &=
    \left( \mathcal{J} - \rho^U  - \rho^V\right) \mathscr{G}
    - \mathscr{G} \left( \mathcal{J} - \rho^U - \rho^V\right) \\
    & + \Delta^U\mathscr{F}^* - \mathscr{F}\Delta^U
\end{split}
\end{equation}
\begin{equation}
\label{eq:dFdt}
\begin{split}
    -i\dfrac{d}{dt} \mathscr{F}
    &=
    \left( \mathcal{J} - \rho^U - \rho^V\right) \mathscr{F}
    + \mathscr{F} \left( \mathcal{J}^* - \rho^U - \rho^V + 2\mu I \right) \\
    &+ \left( \mathscr{G} -  I \right) \Delta^U - \trl\left( \mathscr{G}\Delta^U \right),
\end{split}
\end{equation}
where the matrix elements are given by  
$\mathcal{J}_{i\sigma ; j\sigma'}$ $=$ $\delta_{\sigma, \sigma'} \mathcal{J}_{ij}$, 
$\rho^U_{i\sigma ; j\sigma'}$ $=$ $U\delta_{i, j} \delta_{\sigma, \sigma'} \mathscr{G}_{i\bar{\sigma} ; i\bar{\sigma}}$, 
$\Delta^{U}_{i\sigma ; j\sigma'}$ $=$ $U \delta_{i, j} (i\sigma_y)_{\sigma,\sigma^{\prime}} \mathscr{F}_{i\uparrow, i\downarrow}$, 
and 
$\rho^V_{i\sigma;j\sigma'}$ $=$ $\delta_{i,j}\delta_{\sigma,\sigma'}
\sum_{k\sigma}\mathscr{G}_{k\sigma;k\sigma}V_{kj}$.
The dimension of each matrix is $2N \times 2N$.
A similar formalism is derived in Refs.~\onlinecite{Chern2019_EoM, Seibold2021_Higgs_DisSC}.

We also define a time-dependent distribution function to track the time evolution of the electronic structure.
The Hamiltonian at time $t$ can be diagonalized as
$\vec{C}^{\dagger} H_{\mathrm{BdG}}(t)\vec{C}$ $=$ $\vec{B}^\dagger(t) D(t) \vec{B}(t)$,
where
%and $\Vec{B}(t)$ is an annihilation operator of the quasiparticle defined as $\Vec{B}(t)=\hat{U}^\dagger (t) \vec{C}$
%and
%$\hat{D}(t)=\mathrm{diag}(E_1,\dots,E_{4N})$.
$\Vec{B}(t)$ $=$ $U^\dagger (t) \vec{C}$, and
$D(t)$ $=$ $\mathrm{diag}(E_1(t)$, $\cdots$, $E_{2N}(t))$.
The $\mu$-th component of $\Vec{B}(t)$, given by
$\Vec{B}_{\mu}(t)= \sum_j \left[ U(t) \right]_{j \mu}^* \vec{C}_{j}$, stands for the annihilation operator of the quasiparticle with eigenenergy $E_\mu(t)$
%The each component of $\Vec{B}(t)$ is given by $\Vec{B}_{\mu}(t)= \sum_j \left[ \hat{U}(t) \right]_{j \mu}^* \vec{C}_{j}$.
The time-dependent distribution function for each $E_\mu(t)$, denoted by $\mathcal{N}_{\mu}(t)$, is calculated as 
\begin{equation}
\begin{split}
    \mathcal{N}_{\mu}(t) 
    &=
    \Braket{\Vec{B}_{\mu}^\dagger(t) \Vec{B}_{\mu}(t)}\\
    &=
    \sum_{i,j} \left[\hat{U}(t)\right]_{i \mu} \left[\hat{U}(t)\right]_{j \mu}^*
    \Braket{\vec{C}_{i}^\dagger (t) \vec{C}_{j} (t)},
\end{split}
\end{equation}
where $\braket{\vec{C}_{i}^\dagger (t) \vec{C}_{j} (t)}$ is calculated from $\mathscr{G}(t)$ and $\mathscr{F}(t)$.
In addition, the time-dependent pair-potential for each $E_\mu(t)$, denoted by $\Delta_\mu(t)$, is calculated as 
\begin{equation}
\begin{split}
    \Delta_{\mu}(\bm{r}_i, t)
    &=
    \left[\hat{U}(t)\right]_{i\mu} \left[\hat{U}(t)\right]_{i+N\mu}^* \Braket{B_\mu^\dagger(t) B_\mu(t)}
\end{split} \label{eq:t.dep.pair.pot}
\end{equation}
This quantity satisfies the following relation: $\sum_\mu \Delta_\mu(\bm{r}_i, t) = \braket{c_{i\downarrow} (t)c_{i\uparrow}(t)}$.

\subsection{Linear response theory}
The charge and pair correlation functions for an imaginary time $\tau$ are, respectively, defined as
\begin{align}
    \Pi_{\textrm{c}}(\bm{q}, \bm{q}', \tau)
    &=
    \sum_{\sigma\sigma'} \Pi_{\mathrm{c},\sigma\sigma'}(\bm{q}, \bm{q}', \tau) ,\label{eq:corr.c.time1}\\
    \Pi_{\mathrm{c},\sigma\sigma'}(\bm{q}, \bm{q}', \tau) 
    &=
    -\dfrac{1}{N}\Braket{ T_\tau \rho_{\bm{q}\sigma}(\tau) \rho_{-\bm{q}'\sigma'}(0) } \label{eq:corr.c.time2},\\
    \Pi_{\textrm{SC}}(\bm{q}, \bm{q}', \tau)
    &=
    -\dfrac{1}{N}
    \Braket{ T_\tau \Delta_{\bm{q}}(\tau) \Delta_{\bm{q}'}^\dagger(0) },\label{eq:corr.sc.time}
\end{align}
where $\rho_{\bm{q}\sigma}=\sum_{\bm{k}} c_{\bm{k}\sigma}^\dagger c_{\bm{k}+\bm{q}\sigma}$, $\Delta_{\bm{q}}=\sum_{\bm{k}} c_{-(\bm{k}+\bm{q})\downarrow} c_{\bm{k}\uparrow}$, and $T_{\tau}$ is the imaginary-time ordered product. Their Fourier transform in the frequency domain is given with the bosonic Matsubara frequency $\varepsilon_\ell = 2\pi \ell T$ by 
\begin{align}
    \Pi_{\textrm{c(SC)}}(\bm{q}, \bm{q}', i\epsilon_\ell)
    &=
    \int_0^{1/T} d\tau e^{i\epsilon_\ell\tau}
    \Pi_{\textrm{c(SC)}}(\bm{q}, \bm{q}',\tau). \label{eq:corr.func.}
\end{align}
The Fourier transform of Eq.~\eqref{eq:corr.c.time2} is defined similarly.
 We calculate these correlation functions within the RPA. The RPA takes into account the self-consistent dynamics of the mean fields due to the applied external fields. 
 We formulate the correlation functions within the RPA, by following Refs.~\onlinecite{Schrieffer1989, Rowe2012}. 

In the following, we construct the RPA formalism in the presence of the CDW order characterized by the order vector $\bm{Q} = (\pi, \pi)$. 
First we should remark on the momenta in the argument of  $\Pi_{\mathrm{c(SC)}}$ in Eq.~\eqref{eq:corr.func.}. To analyze the excitation structure, we are interested in the diagonal elements $\Pi_{\mathrm{c(SC)}}(\bm{q},\bm{q}, i \epsilon_\ell)$. However, the 
$(\bm{q}, \bm{q} + \bm{Q})$ component 
is also taken into account through the intermediate process of the RPA, as can be seen below. For this purpose, we introduce a 4 by 4 matrix $\hat{\Pi}_{\mathrm{c},\bm{q}}$
and a 2 by 2 matrix
$\check{\Pi}_{\mathrm{SC},\bm{q}}$
defined as
$[\hat{\Pi}_{\mathrm{c},\bm{q}}]_{\bm{q}_1\sigma_1:\bm{q}_2\sigma_2} = \Pi_{\mathrm{c},\sigma_1\sigma_2}(\bm{q}_1,\bm{q}_2,i\epsilon_\ell)$
and 
$[\check{\Pi}_{\mathrm{SC},\bm{q}}]_{\bm{q}_1:\bm{q}_2} = \Pi_{\mathrm{SC}}(\bm{q}_1,\bm{q}_2,i\epsilon_\ell)$, respectively, where $\bm{q}_{1,2}$ takes either $\bm{q}$ or $\bm{q}+\bm{Q}\equiv \bar{\bm{q}}$.
The definitions given by Eq.~\eqref{eq:corr.func.} are convenient for the following matrix RPA form.
In the presence of the CDW order, the following Green's functions take nonzero values: 
$G_{\bm{k}\sigma}(\tau) = -\braket{T_\tau\, c_{\bm{k}\sigma}(\tau)c_{\bm{k}\sigma}^\dagger(0)} =: G_{\bm{k}}(\tau)$ and
$D_{\bm{k}\sigma}(\tau)=-\braket{T_\tau\, c_{\bm{k}\sigma}(\tau) c_{\bm{k}+\bm{Q}\sigma}^\dagger(0) }=:D_{\bm{k}}(\tau)$, where they are  independent of the spin index because the SDW order is neglected.
Within the RPA, correlation functions in the matrix form are given by
\begin{align}
\label{eq:RPA-c}
    \hat{\Pi}_{\textrm{c},\bm{q}}^{\mathrm{RPA}}(i\epsilon_\ell)
    &=
    \left\{ \hat{I} - \hat{\Pi}_{\textrm{c},\bm{q}}^0(i\epsilon_\ell) \cdot \hat{U}_{\textrm{c},\bm{q}}\right\}^{-1} \cdot \hat{\Pi}_{\textrm{c},\bm{q}}^0(i\epsilon_\ell),\\
\label{eq:RPA-sc}
    \check{\Pi}_{\textrm{SC},\bm{q}}^{\mathrm{RPA}}(i\epsilon_\ell)
    &=
    \left\{ \check{I} - \check{\Pi}_{\textrm{SC},\bm{q}}^0(i\epsilon_\ell) \check{U}_{\textrm{SC},\bm{q}} \right\}^{-1}  \check{\Pi}_{\textrm{SC},\bm{q}}^0(i\epsilon_\ell),
\end{align}
where $\hat{I}(\check{I})$ denotes the 4 by 4 (2 by 2) identity matrix, and $\hat{U}_{\mathrm{c},\bm{q}}$ and $\check{U}_{\mathrm{SC},\bm{q}}$ are the interaction matrices defined for the basis sets $(\bm{q}\uparrow, \bm{q}\downarrow, \bar{\bm{q}}\uparrow, \bar{\bm{q}}\downarrow)$ and $(\bm{q},\bar{\bm{q}})$, respectively.
Their matrix elements are, respectively, defined as
\begin{align}
    \hat{U}_{\mathrm{c},\bm{q}} &=
    \begin{pmatrix}
        V_{\bm{q}} & U + V_{\bm{q}} & 0 & 0 \\
        U + V_{\bm{q}} & V_{\bm{q}} & 0 & 0 \\
        0 & 0 & -V_{\bm{q}} & U - V_{\bm{q}} \\
        0 & 0 & U - V_{\bm{q}} & -V_{\bm{q}}
    \end{pmatrix},\\
    \check{U}_{\mathrm{SC},\bm{q}} &=
    \begin{pmatrix}
        U + V_{\bm{q}} & 0\\
        0 & U - V_{\bm{q}}
    \end{pmatrix}.
\end{align}
In Eqs.~\eqref{eq:RPA-c} and \eqref{eq:RPA-sc},  
$\hat{\Pi}_{\mathrm{c}}^{0}$ and $\check{\Pi}_{\mathrm{sc}}^{0}$
are the lowest-order correlation functions that include the Hartree--Fock contributions in the single particle Green's functions. By noting that  $\Pi^0_{\sigma\sigma'}(\bm{q}, \bm{q}', i\epsilon_\ell)=\delta_{\sigma,\sigma'}\Pi_{\textrm{c}}^0(\bm{q}, \bm{q}', i\epsilon_\ell)$ and 
$\Pi_{\mathrm{c,SC}}^0(\bm{q}_1,\bm{q}_2,i\epsilon_\ell) = \Pi_{\mathrm{c,SC}}^0(\bm{q}_2,\bm{q}_1,i\epsilon_\ell)$, the independent components can be explicitly written as 
\begin{align}
     \Pi_{\textrm{c}}^0(\bm{q}, \bm{q}, i\epsilon_\ell)
    &=
    \dfrac{T}{N}\sum_{\bm{k}, \ell} \left\{ 
    G_{\bm{k}}(i\omega_n) G_{\bm{k} + \bm{q}}(i\epsilon_\ell + i\omega_n)
    \right.\nonumber \\
    &\hspace{1em}\left.
    + D_{\bm{k}}(i\omega_n)^\dagger D_{\bm{k}+\bm{q}}( i\epsilon_\ell + i\omega_n)
    \right\}, \\
    \Pi_{\textrm{c}}^0(\bm{q}, \bar{\bm{q}}, i\epsilon_\ell)
    &=
    \dfrac{T}{N}\sum_{\bm{k}, \ell} \left\{ 
    G_{\bm{k}}(i\omega_n) D_{\bm{k} + \bm{q}}(i\epsilon_\ell + i\omega_n)
    \right.\nonumber \\
    &\hspace{1em}\left.
    + D_{\bm{k}}^\dagger(i\omega_n) G_{\bm{k} + \bm{q}}(i\epsilon_\ell + i\omega_n)
    \right\}, \\
    \Pi_{\textrm{SC}}^0(\bm{q}, \bm{q}, i\epsilon_\ell)
    &=
    -\dfrac{T}{N}\sum_{\bm{k}, \ell} \left\{ 
    G_{\bm{k}}(-i\omega_n) G_{-(\bm{k} + \bm{q})}(i\epsilon_\ell + i\omega_n)
    \right.\nonumber \\
    &\hspace{1em}\left.
    + D_{\bm{k}}(-i\omega_n) D_{-(\bm{k}+\bm{q})}(i\epsilon_\ell + i\omega_n)
    \right\}, \\
    \Pi_{\textrm{SC}}^0(\bm{q}, \bar{\bm{q}}, i\epsilon_\ell)
    &=
    -\dfrac{T}{N}\sum_{\bm{k}, \ell} \left\{ 
    G_{\bm{k}}(-i\omega_n) D_{-(\bm{k} + \bm{q})}(i\epsilon_\ell + i\omega_n)
    \right.\nonumber \\
    &\hspace{1em}\left.
    + D_{\bm{k}}(-i\omega_n) G_{-(\bm{k}+\bm{q})}(i\epsilon_\ell + i\omega_n)
    \right\}.
\end{align}
To investigate the excitation structure, we perform an analytic continuation of  
$\sum_{\sigma,\sigma^{\prime}}[\hat{\Pi}_{\mathrm{c},\bm{q}}^{\mathrm{RPA}}(i\epsilon_\ell)]_{\bm{q}\sigma:\bm{q}\sigma^{\prime}}$ and 
$[\check{\Pi}_{\mathrm{SC},\bm{q}}^{\mathrm{RPA}}(i\epsilon_\ell)]_{\bm{q}:\bm{q}}$: $i\epsilon_\ell \to \omega + i\delta$, and describe them as  $\Pi_{\mathrm{c},\bm{q}}^{\mathrm{RPA}}(\omega)$
and $\Pi_{\mathrm{SC},\bm{q}}^{\mathrm{RPA}}(\omega)$, respectively.

\section{RESULTS}
\subsection{Setup}
\begin{figure}[t]
    \centering
    \includegraphics[width=1.0\linewidth]{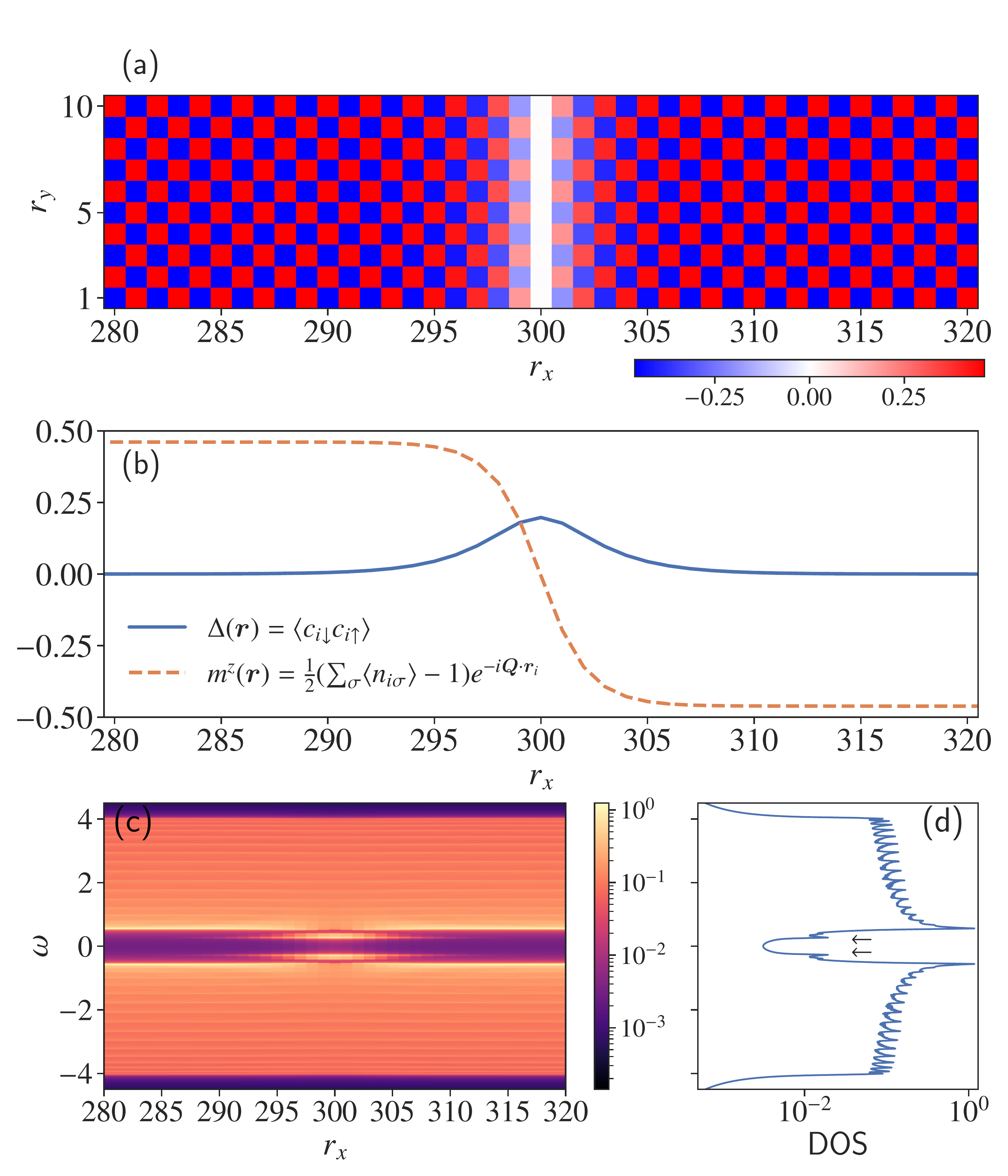}
    \caption{
    (a) Charge density distribution in real space.
    An interface is located at $r_x=300$ in the system of size $401\times40$.
    (b) Order parameters of SC $\Delta(\bm{r})$ and CDW $m^z(\bm{r})$ along the $x$ direction.
    The interface is dominated by $\Delta(\bm{r})$ rather than $m^z(\bm{r})$, while in the region far from the interface $\Delta(\bm{r})$ goes to 0 and $m^z(\bm{r})$ is dominant. 
    (c) Local density of states (LDOS) around the domain wall.
    The horizontal axis denotes the spatial direction across the domain wall ($r_x$), while the vertical axis denotes the energy ($\omega$). along the $x$ direction. The intensities of the LDOS are indicated by the color map.
    (d) Density of states plotted on a logarithmic scale. Arrows indicate the DWBSs. The vertical axis is the same as that in (c).
    }
    \label{fig:setup}
\end{figure}

Before showing numerical results based on the above formulation, we summarize the parameters and numerical conditions. As a unit of energy, we set $|J| = 1$.
In the following numerical results, we use $U=-2$, $V=0.05$.
The total number of electrons $N_{\mathrm{e}}$ is fixed to $N$, corresponding to the half-filling case. 
A non-zero $V$ lifts the degeneracy between the CDW state and the SC state, and the CDW state is realized as the uniform ground state in the half-filling case.

As an initial state of the time evolution, we consider two CDW domains with opposite signs, which induce the superconductivity along their interface \cite{Leridon2020}.
The geometry of the system is an $N_x \times N_y = 401 \times 40$ site lattice with periodic boundary conditions (PBCs) in both directions.
When $N_x$ is odd, the PBC in the $x$ direction naturally introduces an interface along the $y$ direction as a self-consistent solution. 
Here, we use the following integer notations $i = (i_x, i_y)=\bm{r}_i$ and $\bm{r} = (r_x, r_y)$ interchangeably to represent a site position.
Figure~\ref{fig:setup}(a) shows the spatial profile of the charge density $\sum_{\sigma}(\braket{c_{i\sigma}^\dagger c_{i\sigma}} - 1)$.
Figure~\ref{fig:setup}(b) shows the $r_x$ dependence of the SC order parameter $\Delta(\bm{r}) = \braket{c_{i\downarrow} c_{i\uparrow}}$ using the dashed line, and that of the staggered density $m^z(\bm{r})=(\sum_\sigma\braket{n_{i\sigma}} - 1)e^{i\bm{Q}\cdot\bm{r}_i}/2$ using the solid line.
Note that $\Delta(\bm{r})$ and $m^z(\bm{r})$ are uniform along the $y$ direction. 
At the center of the domain wall $r_x=300$, $\Delta_i$ reaches a maximum value, and $m^z$ is zero.
%\red{Their definitions are, respectively, given by
%$m(r_x) = m(r_x, r_y) = n(\bm{r}_j = (r_x, r_y))e^{i \bm{Q}\cdot\bm{r}_j}$ and $\Delta(r_x) = \Delta(r_x, r_y) = \Delta(\bm{r}_j)$, which are independent of $r_y$ in equilibrium for half filling.} 

In Figs.~\ref{fig:setup}(c) and \ref{fig:setup}(d), we plot the local density of states (LDOS) and density of states (DOS), defined by
$\mathrm{LDOS}(r_x, \omega) = -1/(\pi N_{y})\sum_{r_y}\mathrm{Im}\,\mathrm{tr}\, [(\omega + i\eta - H_{\mathrm{BdG}})^{-1}]_{\bm{r},\bm{r}}$ and $\mathrm{DOS}(\omega) = 1/N_x\sum_{r_x} \mathrm{LDOS}(r_x, \omega)$, where $\mathrm{tr}$ is the trace in Nambu space.
The figures show that the system has a CDW gap $\omega_{\varg}\simeq1.1$.
In (c), there are fermionic states bound in the domain wall indicated by the arrows in (d).
In this paper, we call them domain-wall bound states (DWBSs).

Such a non-uniform structure is related to a magnetic domain wall~\cite{Leridon2020}.
The low-energy state of the EAHM are described by the order parameters of the CDW and the superconductivity, which can be regarded as an antiferromagnetic order in a classical pseudospin system. Consider the following map~\cite{Dichtel1971,Shiba1972,Nagaoka1974}: $S_j^x + i S_j^{y} = c_{j\downarrow}c_{j\uparrow}e^{i\bm{Q}\cdot\bm{r}_j}$, $S_j^z = (n_{j\uparrow} + n_{j\downarrow} - 1)/2$, the interaction between nearestneighbor peseudospins are antiferromagnetic. The Ne\'{e}l order along the $z$ direction in pseudospin space corresponds to the CDW, and the Ne\'{e}l order in the $x$-$y$ plane corresponds to uniform superconductivity. 
In the low-energy region, a spatially local order-parameter manifold
is constructed on the SO(3) sphere of $\vec{S}_j e^{i\bm{Q}\cdot\bm{r}_j}$. In the absence of $V$, the ground state has SO(3) symmetry. A small $V (>0)$ lifts this degeneracy, so that the Ne\'{e}l order along the $z$-axis has a lower energy than that in the $x$-$y$ plane; the north (south) pole denotes a positive (negative) value of the CDW order parameter 
$m^z(\bm{r}_j) = \braket{S_j^ze^{i\bm{Q}\cdot\bm{r}_j}} >0\ (m^z(\bm{r}_j) < 0)$.  
The aforementioned structure can be regarded as a pseudospin antiferromagnetic domain. In the domain wall region, antiferromagnetic spin structures with $x$ and $y$ components $\braket{S_j^{x,y}e^{i\bm{Q}\cdot\bm{r}_j}}$ correspond to the uniform SC structure with U(1) phase degrees of freedom, 
$\mathrm{Re}\,\Delta(\bm{r})$, $\mathrm{Im}\,\Delta(\bm{r})$~\cite{Leridon2020}.
The domain-wall width depends on the off-site interaction $V$ and the width diverges as $V \to 0$, because $V$ plays a role of easy-axis anisotropy along the $z$-axis in the pseudospin picture. For the above parameter set, the width along the $x$ direction in real space is estimated to be $10$ sites Fig.~\ref{fig:setup}(b).

The equation of motion is numerically solved using the fourth-order Runge--Kutta method (RK4) \red{for the time-step $\Delta t = 0.01$}. 
%The time step was set as $\Delta t = 0.01$.
By taking advantage of the translational symmetry along the $y$ direction Fig.~\ref{fig:setup}(a), we performed the Fourier transform along the $y$ direction.
%introduce a Fourier transformed electron annihilation operator as 
%$c_{i = (i_x, i_y)} = 1/\sqrt{N_y} \sum_{k_y} e^{ik_y r_y} c_{i_x, k_y}$.
The details are explained in Appendix~\ref{sec:fourier transform}.

In the following subsections, we investigate the photoinduced dynamics of the above interface structure driven by an oscillating electric field with a frequency $\omega_{\mathrm{ext}}$ via the Peierls substitution of the vector potential $\bm{A}$.
The vector potential modifies the hopping amplitude to $\mathcal{J}_{ij} \rightarrow \mathcal{J}_{ij} \exp[-i \bm{A}(t)\cdot (\bm{r}_i - \bm{r}_j)]$.
We choose the vector potential to be $\bm{A}(t)=\bm{e}_{p}A(t) =  \bm{e}_{p}A_0\sin(\omega_{\mathrm{ext}}t)$ with amplitude $A_0$, polarization $\bm{e}_p$, and driving frequency $\omega_{\mathrm{ext}}$. The polarization direction is set as $\bm{e}_p=(1,0)$, where we have checked that the $y$ component of $\bm{e}_p$ does not bring about any noticeable dynamics.

%\subsection{Response for a low frequency}
\subsection{Pseudospin dynamics: $\omega_{\mathrm{ext}} \ll \omega_{\varg}$}
\begin{figure*}
    \centering
    \includegraphics[width=1.0\linewidth]{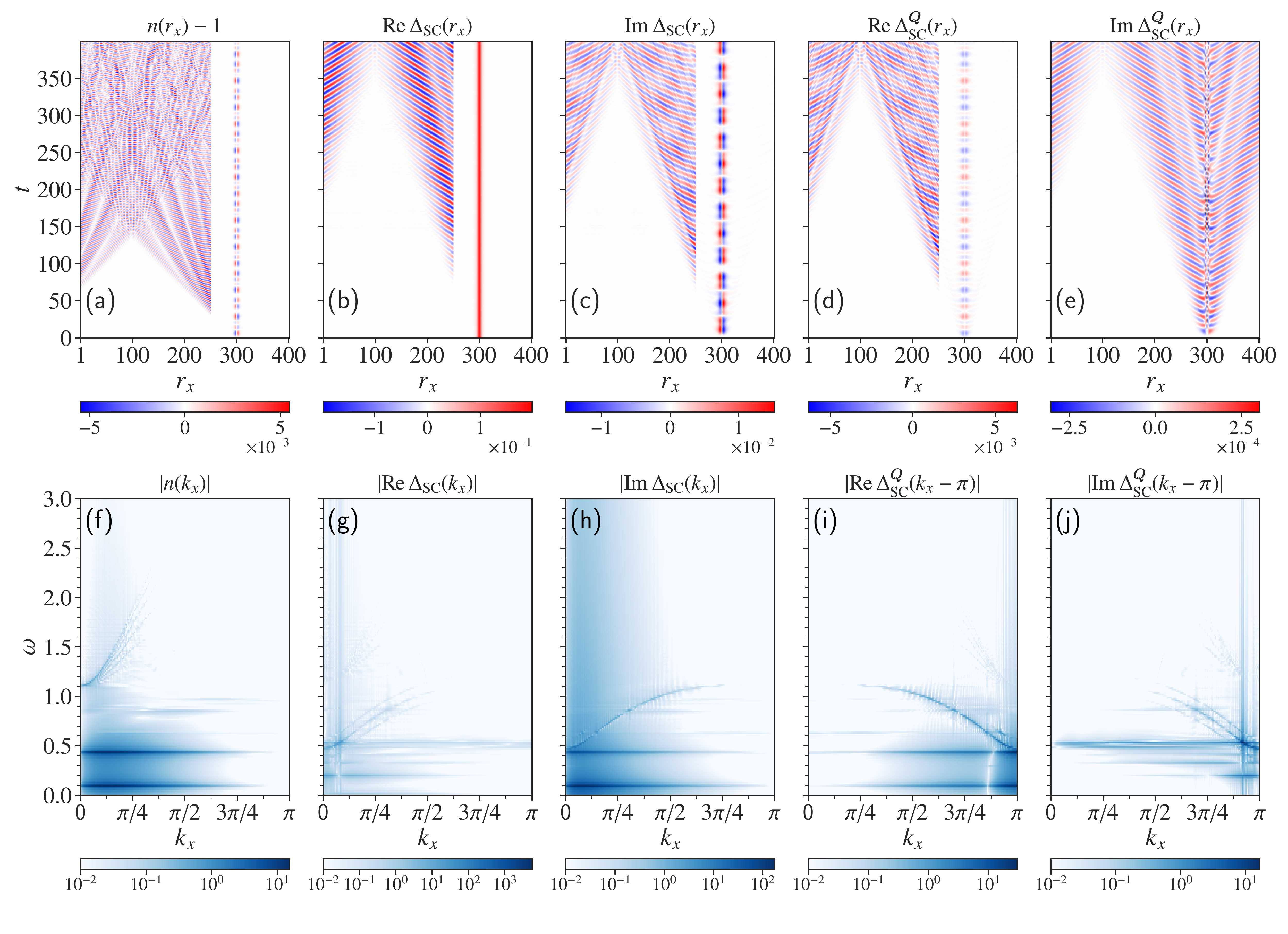}
    \caption{
    (upper panels) Time evolution of order parameters for $A_{0}=0.02$ and $\omega_{\mathrm{ext}}=0.1$.
    (a) Deviation of the charge density from the averaged value.
    (b)(c) Uniform and (d)(e) staggered SC order parameters.
    To visualize the propagation of the collective modes, we multiply the actual values in the range $1 \le r_x \le 250$ by 500 in (a)(b) and 50 in (c)(d).
    (lower panels) Intensities of the Fourier transforms of the corresponding time evolution in real space in the upper row.
    }
    \label{fig:dynamics}
\end{figure*}
In this subsection, we examine the photoinduced dynamics with driving frequency $\omega_{\mathrm{ext}} = 0.1$, which is lower than the CDW gap $\omega_{\varg}\simeq1.1$. In this regime, it is expected that hardly any quasiparticles are excited and the pseudospin picture is valid. The amplitude $A_0$ is set to $0.02$. 

The top panels in Fig.~\ref{fig:dynamics} show the real-space and real-time evolutions of the mean field $O(r_x,t) \equiv {N_y}^{-1}\sum_{r_y} O (\bm{r}=(r_x, r_y), t)$, where $O$ in each panel is as follows:
(a) density deviation from the averaged value $n(\bm{r}) - 1$,
(b) (c) the real and imaginary parts of the uniform SC order parameter $\Delta_{\mathrm{SC}}(\bm{r})\equiv\Delta(\bm{r})$,
and (d) (e) the real and imaginary parts of the staggered SC order parameter $\Delta_{\mathrm{SC}}^{Q}(\bm{r}) \equiv \Delta(\bm{r})e^{i\bm{Q}\cdot\bm{r}}$. 
In panels (b), the SC interface, initially at $r_x = 300$, hardly changes its position and the SC phase during the time evolution. The other panels, (a) and (c)--(e), show the polarization of the mean fields in the interface, which represents a deformation of the domain wall. 

It should be also noted that the uniform and staggered SC components, propagate from the domain wall, as can be seen in panels (b)--(f), i.e., the collective mode propagates. 
%This can be interpreted as the propagation of the collective modes in the abovementioned pseudospin picture. 
% pseudospins away from the interface are oriented in the two opposite pseudospin directions representing the positive and negative CDW order parameters.
The laser irradiation drives the deformation of the domain wall. 
Such an internal deformation generates a pseudospin wave propagating outwards. In the region far from the domain wall, the pseudospin wave can be regarded as precessional motion of pseudospins around the $z$-axis corresponding to the CDW, i.e., excitation of the $x$, $y$ components corresponding to the SC components.
The precession propagates from the interface to the outside, which can be interpreted as the pseudospin wave emission from the domain walls.
This pseudospin wave contains the uniform and staggered components as in the case of the low-energy antiferromagnetic spin waves. 
%In more details, this pseudospin wave has an analogy of the antiferromagentic spin waves because the uniform and the staggered components exist owing to the bipartite structure.
It is known that spin waves can also be emitted from a domain wall in the ferromagnetic case and the antiferromagnetic case by using oscillating magnetic fields or spin orbit torques~\cite{Oh2017,Shiino2016}. The pseudospin-wave emission in the present study is an analogous to these magnetic systems, but the emission is triggered by the electric field. (As mentioned in Sect.~III~A, the domain wall can be interpreted as an antiferromagnetic domain wall with easy-axis anisotropy along the $z$-axis.)
In terms of the collective excitation of the SC order parameter, 
the propagating wave is a phase rotation of $\Delta_{\mathrm{SC}}$ with $\bm{k}\sim (0,0)$ and $(\pi,\pi)$. 
Note that the staggered component in this case can also be regarded as the $\eta$ pairing excitation of the attractive Hubbard model~\cite{Yang1989,Yang1990,Eugene1996,Kitamura2016}.
The mass of the phase mode is due to the off-site Coulomb interaction $V$, whose effects on the collective mode are discussed at the end of this subsection.

Next, to investigate the emitted collective mode in the uniform region, we perform a Fourier transform into momentum and frequency spaces:
\begin{align}
O(k_x,\omega) = N_y\sum_{r_x} \int_0^{T_{\mathrm{max}}} dt\, e^{i(k_x r_x - \omega t)} O(r_x, t),
\end{align}
where $T_{\mathrm{max}}=400$.
The absolute values of the Fourier components are shown in Figs.~\ref{fig:dynamics}(f)--\ref{fig:dynamics}(j).
By applying a laser with zero momentum, the \red{spectral intensities} can be acquired not only at $\bm{k}=0$ but also in the finite $\bm{k}$ region owing to the inhomogeneity of the domain wall. 
The flat peaks in $\omega=\omega_{\mathrm{ext}}=0.1$ in each panel correspond to the forced oscillation by the external electric field.
Figures~\ref{fig:dynamics}(g) and \ref{fig:dynamics}(h) [\ref{fig:dynamics}(i) and \ref{fig:dynamics}(j)] show the dispersive SC collective modes for $\omega \lesssim \omega_{\varg}$ and $0 \le k_x \le \pi$ with a uniform (staggered) profile along the $y$ direction.
Note that 
using $\Delta(k_x, k_y,\omega) = \sum_{\bm{r}}\int_0^{T_{\max}}dt e^{i(\bm{k}\cdot\bm{r} - \omega t)} \Delta(\bm{r})$, $\Delta_{\mathrm{SC}}(k_x)= \Delta(k_x, k_y = 0)
$ and 
$\Delta_{\mathrm{SC}}^Q(k_x-\pi)= \Delta(k_x, k_y = \pi)
$.
As a result of the folding of the first Brillouin zone by the CDW order, the induced spectral intensities concentrate around $\bm{k}=(0,0)$ and $\bm{k}=(\pi, \pi)$.
In addition, Floquet side bands surround the SC collective mode with energy difference $\pm \omega_{\mathrm{ext}}$.
We have checked that the Floquet side band exists at the other driving frequencies $\omega_{\mathrm{ext}}$.
Those dispersion can be interpreted as a photon-dressed collective mode.
A similar dispersive branch is observed for $\omega \gtrsim \omega_{\varg}$ in Fig.~\ref{fig:dynamics}(f). However, this is not a collective mode of the charge density, but rather part of the continuum spectra of the particle-hole excitation, which is shown more clearly in the following analysis.

%\red{In Fig.~\ref{fig:dynamics}~(l), we find two peaks emerge at $\omega=2\omega_{\mathrm{ext}}, \omega_{\mathrm{ext}}+0.5$.}

\begin{figure}[t]
    \centering
    \includegraphics[width=1.0\linewidth]{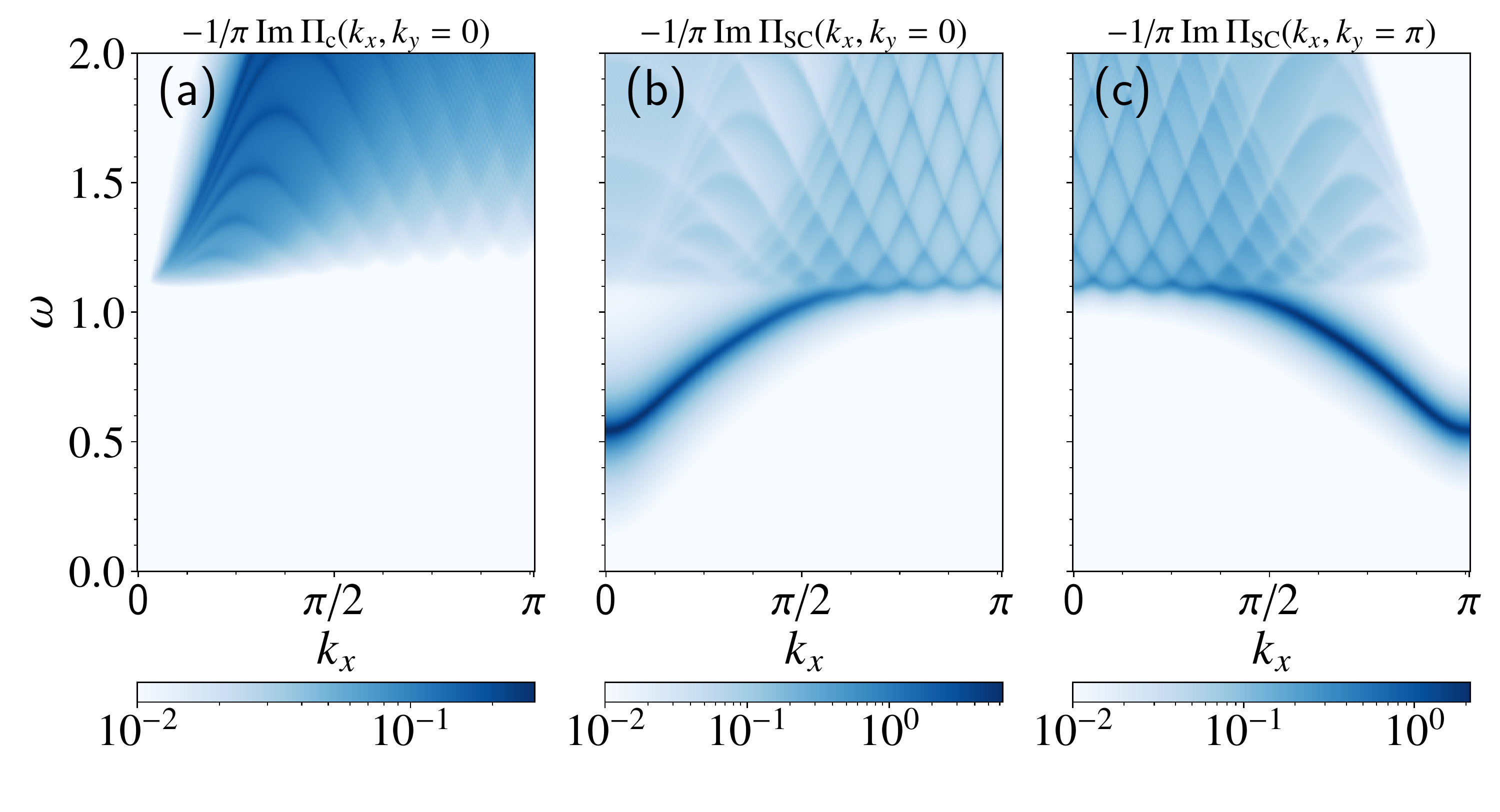}
    \caption{
    Intensity map of the dynamical correlation functions for (a) charge and (b)(c) SC pair potential along the $k_x$ direction.
    The equilibrium state was a uniform CDW order and the system was of size $400 \times 40$. The top panels show the imaginary parts. The value of $k_y$ is shown in each panel.
    }
    \label{fig:susceptibility}
\end{figure}
In order to analyze the collective modes emitted from the domain wall, we calculated the dynamical charge and pair correlation functions based on the RPA (Eqs.~\eqref{eq:RPA-c} and \eqref{eq:RPA-sc}).
Figures~\ref{fig:susceptibility} show the correlation functions as a function of $k_x$ for $k_y=0$ and those for $k_y=\pi$.
%The imaginary parts of the dynamical charge correlation function [Fig.~\ref{fig:susceptibility}(a)] and pair correlation functions of the uniform SC [Fig.~\ref{fig:susceptibility}(b)] and the staggered SC [Fig.~\ref{fig:susceptibility}(c)] show good agreements with the results obtained from the real-space and real-time calculations shown in Fig.~\ref{fig:dynamics}(f), Figs.~\ref{fig:dynamics}(g-h), and Figs.~\ref{fig:dynamics}(i-j), respectively.
The imaginary parts of the dynamical correlation functions in Fig.~\ref{fig:susceptibility} are in good agreement with the results obtained from the real-space and real-time calculations in Fig.~\ref{fig:dynamics} [the charge density:~Fig.~\ref{fig:susceptibility}(a) and Fig.~\ref{fig:dynamics}(f), 
the uniform SC pair:~Fig.~\ref{fig:susceptibility}(b) and Figs.~\ref{fig:dynamics}(g), \ref{fig:dynamics}(h), 
and the staggered SC pair:~Fig.~\ref{fig:susceptibility}(c) and Figs.~\ref{fig:dynamics}(g), \ref{fig:dynamics}(h)].
Figure~\ref{fig:susceptibility}(a) shows continuum spectra rather than a collective mode. 

By contrast, Figs.~\ref{fig:susceptibility}(b) and \ref{fig:susceptibility}(c) show a collective mode with an energy below the continuum spectra. 
When the inter-site Coulomb interaction $V$ is zero, because of the SO(3) symmetry of the pseudospin, gapless NG modes appear at $\bm{q} = (0,0)$ and $(\pi,\pi)$~\cite{Kostyrko1992, Hoshino2014, Cea2015_CMinAHM}, just as in the antiferromagnetic case with the Ne\'{e}l order.
In the presence of a repulsive $V ( > 0)$, such a collective mode requires a finite excitation energy.
In Figs.~\ref{fig:susceptibility}(b) and \ref{fig:susceptibility}(c), the collective modes below the continuum are the massive Nambu--Goldstone (NG) modes, where the term ``NG mode'' refers due to the propagation of phase rotation of the SC mean fields.
%We also plot the real part of correlation functions in Fig.\ref{fig:dynamics}[*].
%These spectra show the range in which the forced oscillations occur.
%In the region above $\omega \geq 1$, 
%\blue{we can see that the quasiparticle continuum is almost not excited} \red{(lower panels of Fig.~\ref{fig:dynamics})}. 
The lower panels of Fig.~\ref{fig:dynamics} show very small spectral intensities in the region $\omega \geq 1$, which indicates that the external field does not excite quasiparticles in the continuum-spectral region.
This is why the pseudospin description works quite well.
%We should note that this calculation based on linear response theory did not include any side bands around the collective mode, but that Floquet theory would include the side band around the SC collective mode.
We should note that this calculation based on linear response theory cannot reveal any side bands around the collective mode, whereas that Floquet theory could be used to reveal the side band around the SC collective mode.

We should also note that the energies of the SC collective mode obtained by the RPA are slightly higher than those observed in the real-space calculation. This may have been because we neglected vertex-type diagrams due to $V$ for simplicity.

%\subsection{Response for a high frequency}
\subsection{Quasiparticle excitation: $\omega_{\mathrm{ext}} \gtrsim \omega_{\varg}$}
\begin{figure}[t]
    \centering
    \includegraphics[width=1.0\linewidth]{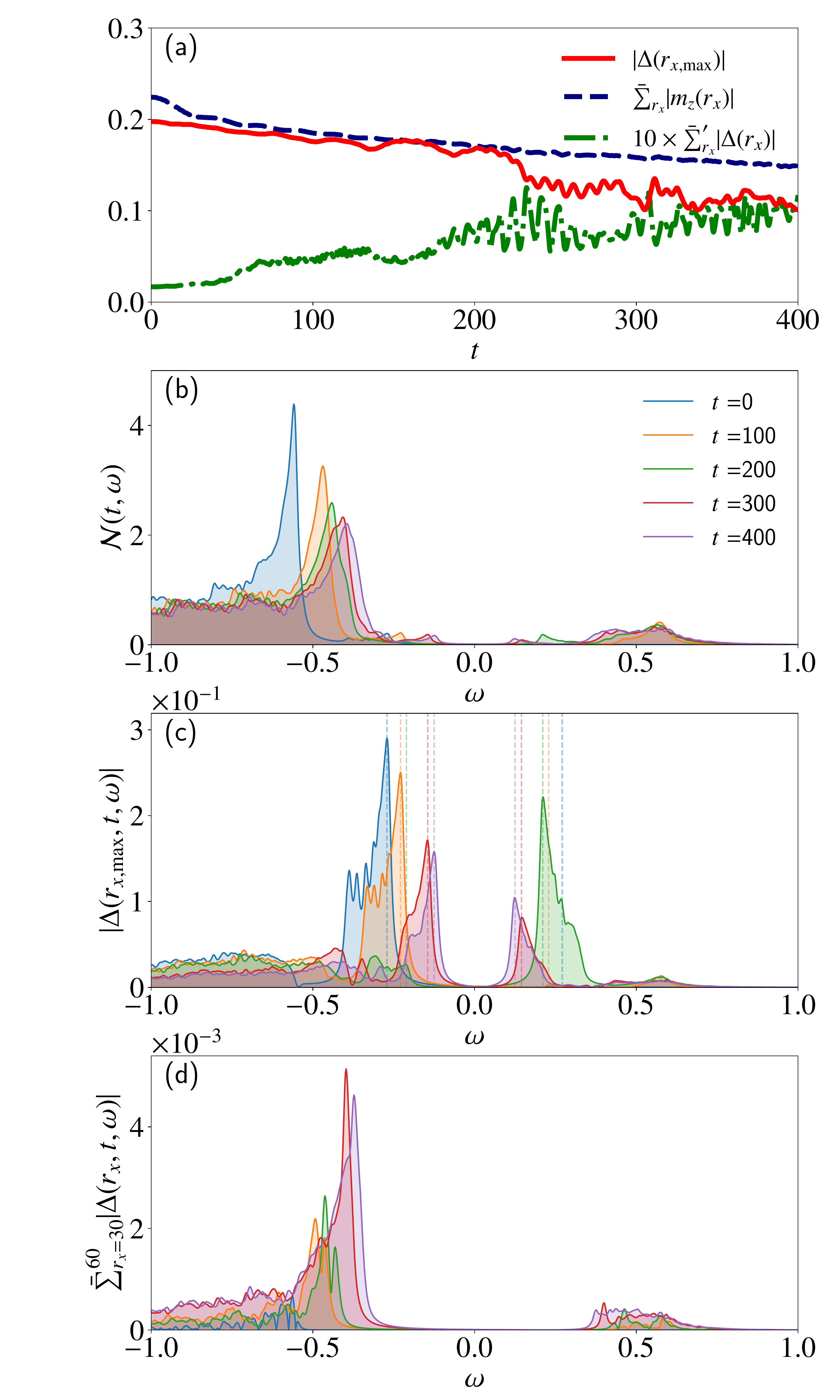}
    \caption{
    (a) Time evolution of the maximum value of $|\Delta(r_x)|$ (solid), the averaged CDW order parameter $m^z$ (dashed), and the averaged SC order parameter in the region far from the domain wall (dashed-dotted). 
    $|\Delta(r_x)|$ is maximized at the center of the domain wall, denoted by $r_{x,\mathrm{max}}$.
    (b) Time evolution of the quasiparticle population defined by $\mathcal{N}(t, \omega) = \sum_{\mu} \mathcal{N}_{\mu}(t) \delta(\omega - E_\mu)$.
    Time evolution of the energy resolved pair potential at the domain-wall center, $r_x = r_{x,\mathrm{max}}$ (c) and far from the domain wall (d).
    The vertical dotted lines are guides for the eye indicating the energies of the largest peaks for the DWBSs and their counterpart with the opposite signs. 
    We define $|\Delta(r_x, t, \omega)|=\sum_{\mu} |\Delta_{\mu}(r_{x},t)| \delta(\omega - E_\mu)$.
    The broadening factor of the $\delta$-function in (b)--(d) is set to 0.01.
    We have introduced $\bar{\sum}$ to represent the average per site, i.e., the summation divided by the number of terms in $\sum$. 
    We use $\bar{\sum}_{r_x}^\prime$ to denote exclusion of the domain wall region $85 \le r_x \le 95$ when taking the average.
    %Also $\bar{\sum}_{r_x}^{\prime\prime}$ in (d) represents the contributions far from the domain wall $30 \le r_x \le 60$.
    % For the dashed-dotted green line, we exclude the contribution from the domain wall. 
    %Note that the center of the domain wall is located at $r_x=90$.
    }
    \label{fig:Melting}
\end{figure}
In this subsection, we examine the nonequilibrium dynamics at a frequency $\omega=1.2$ near the CDW gap, for which the quasiparticle responses as well as the collective response of the pseudospins are important. 
%\red{In particular, we use two frequencies $\omega_{\mathrm{ext}}=1.2$ (near resonance) and $\omega_{\mathrm{ext}}=1.5$ (above the resonance). before revision?} 
The amplitude is $A_0 = 0.02$, and the system size is $121\times 40$. 

Figure~\ref{fig:Melting}(a) shows suppression of the CDW order and implies a melting of the SC domain wall. The dashed line shows the spatial average of the absolute value of the CDW order parameter, which decreases with time. In this subsection, we take $\Delta(r_x)$
to be the $r_y$-averaged SC mean-field instead of $\Delta_{\mathrm{SC}}(r_x)$ for simplicity. 
For $t \lesssim 200$, the SC amplitude on the SC interface, represented by $|\Delta(r_{x,\mathrm{max}})|$, decreases as well. 
Here, $r_{x,\mathrm{max}}$ denotes the position at which $|\Delta(r_x)|$ is a maximum, that is, the domain wall center. 
\red{Furthermore}, 
$|\Delta(r_{x,\mathrm{max}})|$ shows a sudden reduction for $t \sim 200$. 
\red{By contrast, a gradual increase can be seen in t}he spatial average of $|\Delta(r_x)|$ where the domain wall region $85 \le r_x \le 95$ is excluded, \red{as shown by the dashed-dotted line.} %does not show a drastic reduction but rather a gradual increase. 
In other words, 
the SC interface structure, represented by the locally strong SC amplitude, melts at around $t\sim200$, and the SC order parameter extends to the whole system through propagation of the pseudospin wave.
Figure~\ref{fig:Melting}(b) shows the time evolution of the quasiparticle population. 
The line at $t=0$ represents the initial state, that is, the fully occupied states below $\omega=0$.
The lines for $t>0$ show that the laser excitation reduces the CDW gap and increases the quasiparticle population in the conduction band.

The time evolutions of the energy and spatial-resolved pair-potential amplitude are plotted in Figs.~\ref{fig:Melting}(c) and \ref{fig:Melting}(d), where we have defined $|\Delta(r_x, t, \omega)|=\sum_{\mu} |\Delta_{\mu}(r_{x},t)| \delta(\omega - E_\mu)$ for $\Delta_{\mu}(r_x, t)$ Eq.~\eqref{eq:t.dep.pair.pot}.
%The time evolutions of the energy and spatial-resolved pair potential amplitude are plotted in Figs.~\ref{fig:Melting}(c) and \ref{fig:Melting}(d). Here we have defined $|\Delta(r_x, t, \omega)|=\sum_{\mu} |\Delta_{\mu}(r_{x},t)| \delta(\omega - E_\mu)$, where $\Delta(t, \omega )$ is Eq.~\eqref{eq:t.dep.pair.pot}.
First, let us discuss the energy-resolved structure of the domain-wall superconductivity and its time evolution. 
The energy-resolved pair potential at the domain-wall center $r_x = r_{x,\mathrm{max}}$ is shown in Fig.~\ref{fig:Melting}(c).
From the data at $t=0$, the domain wall superconductivity mainly consists of the DWBS at $\omega \simeq -0.27$, which has a sharp and strong peak structure. It also includes a broader, less intense contribution from the continuum states for $\omega  < - \omega_{\varg} / 2 \sim  - 0.55$.
The following changes occur under an external field at the resonant frequency:
(i) 
the absolute value of the energy of the DWBS decreases, as does their intensities; %, as well as the CDW gap $\omega_{\varg}$; 
(ii) the energy distribution of the superconductivity from the DWBSs shifts to the positive side at $t \sim 200$. 
This population inversion is attributed to the phase rotation of the superconductivity inside the domain wall, as discussed in Appendix~\ref{sec:phase mode bound in the DW}.
s\red{At $t \gtrsim 300$, the intensities of the DWBSs at the positive and negative energies are comparable.}
By focusing on the phase of $\Delta(r_{x,\mathrm{max}},t,\omega)$, \red{however,} the two contributions have opposite phase (not shown), which results in $\Delta(r_{x,\mathrm{max}},t)$ having a small amplitude;
(iii) the intensity of the continuum contribution also decreases. 
As a whole, the superconductivity in the domain wall region is reduced by resonant driving. 

\red{Next, let us examine what happens away from the domain wall center. In such a region,} the SC order parameter \red{increases,} as indicated in Fig.~\ref{fig:Melting}(a). This can be seen also in Fig.~\ref{fig:Melting}(d), which shows the energy-resolved amplitude of the pair potential averaged over $30 \le r_x \le 60$. %Compared with 
\red{In contrast to what is shown in} panel (c), 
%the contributions to the superconductivity are from the continuum states $|\omega| > \omega_{\varg} / 2$ 
\red{the continuum states $|\omega| > \omega_{\varg} / 2$ mainly contribute to the superconductivity,} because the DWBSs are well localized around the domain wall. The \red{small} contribution at $t = 0$ \red{accounts for} the \red{exponential decay of the SC interface, in the region away from the domain wall center}. 
Remarkably, for $t \ge  100$ \red{the energy-resolved pair potential} increases particularly near the gap edge of the valence band. This enhancement is more noticeable for $t \gtrsim 300$, which is after the drastic reduction of the SC order at the domain wall. 

In \red{summary, in} the case of the resonant excitation, 
the excitation of the quasiparticles into the conduction band reduces the CDW gap $\omega_{\varg}$ (or the gap edge $\omega_{\varg} / 2$). 
The quasiparticles near the gap edge of the valence band contribute to the formation of uniform superconductivity. 
This behavior which inclues the melting of the domain wall and the appearance of  superconductivity in the bulk region, is triggered by \red{the following two factors:} quasiparticle excitations\red{, which cannot be described by} the pseudospin picture\red{,} and \red{the existence of the domain wall.}

\section{SUMMARY AND DISCUSSION}
We have considered the laser-induced nonequilibrium dynamics of the non-uniform system containing an SC domain wall sandwiched between CDW domains.
We have found two driving-frequency regimes: (i) when the frequency of the driving laser is below the CDW gap, its dynamics conform to the pseudospin picture. We have found that pseudospin waves are emitted from the domain wall; 
(ii) when the frequency of the laser is approximately equal to the CDW gap, excitation of the quasiparticles causes the CDW gap and the SC domain wall to melt.
As a result, we have found that uniform superconductivity is induced in the whole system.

Laser control of superconductivity by utilizing a kind of NG mode was suggested in Ref.~\onlinecite{Sentef2017_AttHubb}.
This mode corresponds to a rotation of a vector in a plane composed of the SC and CDW order parameters, and can be controlled by tuning the frequency of the laser for $V=0$ where the SC and CDW orders are degenerate. 
However, the time scale of this collective rotation is very slow because the zero mode in equilibrium is utilized and the time scale is determined by an amount of \red{$\eta$ pairing which is weakly excited to hold the pseudospin picture}. In addition, we have not found such a collective mode in the presence of $V>0$ even in the uniform case; that is, the conditions under which it appears are restricted.

Even in the presence of $V > 0$, superconductivity may appear as an interface of two opposite CDW domains as proposed in Ref.~\onlinecite{Leridon2020}.
In this case, the oscillatory dynamics may have an analogy with that of the domain wall and the spin wave in the antiferromagnetic systems. We have clarified that the emission of the collective mode is possible in a uniformly oscillating electric field owing to the non-uniformity of the domain wall. We have also proposed a possibility of a kind of photoinduced uniform superconductivity via melting of the domain wall and the CDW order that is caused by the resonant excitation of the quasiparticles.
These results suggest that the SC and CDW orders can be controlled by resonant laser excitation in a non-uniform CDW system.
%Especially, when we consider a system in which coherent CDW islands are distributed, the SC appears in a interface between two CDW islands.
%In such a case, our findings suggest the SC can be induced by the laser irradiation.

\begin{acknowledgments}
This work was supported by JST, the establishment of university fellowships
towards the creation of science technology innovation, Grant Number JPMJFS2102 and JSPS KAKENHI, Grants Nos. JP19K14662 and JP22H01221. H.M. is supported by KAKENHI grant Nos. 21H04446, 21H03455, 21K03380, and 20K03769, and by CSIS, Tohoku University.
\end{acknowledgments}

\appendix
\section{Fourier transform in the $y$ direction} \label{sec:fourier transform}
\begin{figure*}
    \centering
    \includegraphics[width=1.0\linewidth]{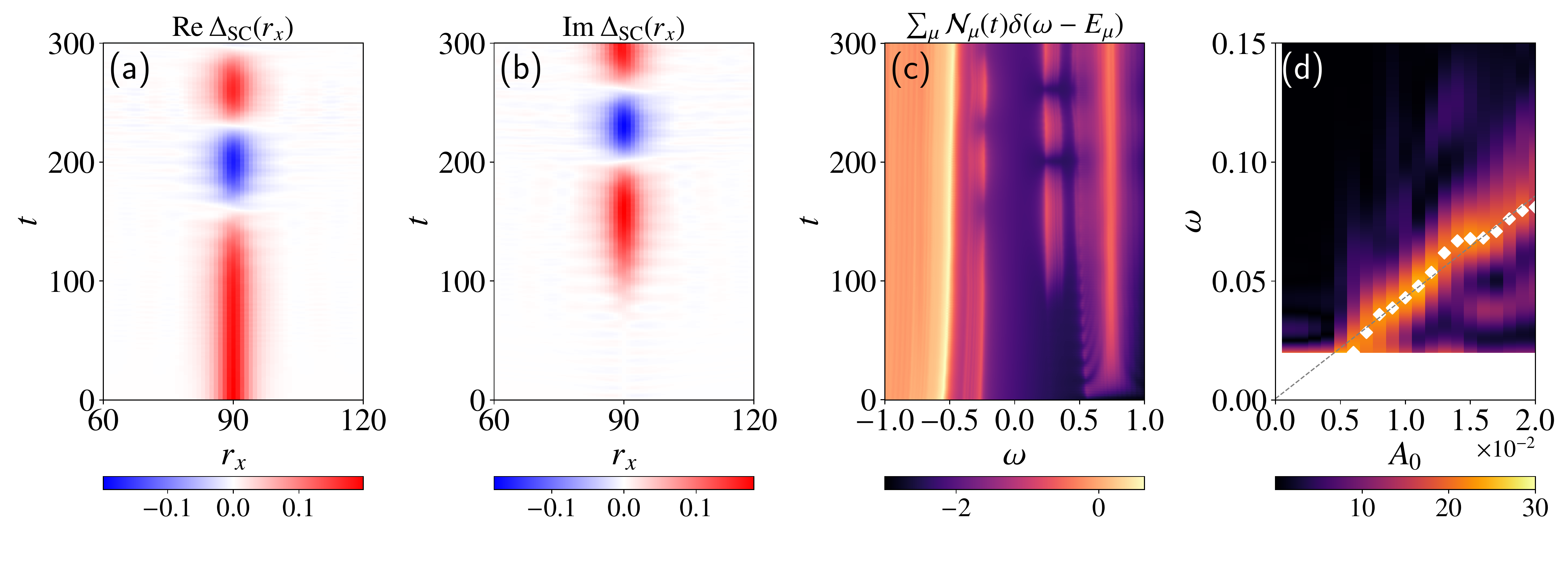}
    \caption{
    Time evolution of the SC order parameter for  (a) the real part and (b) the imaginary part.
    (c) Time evolution of the distribution function represented by $\sum_{\mu} \mathcal{N}_{\mu}(t) \delta(\omega - E_\mu)$ plotted on a logarithmic scale.
    The broadening factor of the $\delta$-function is set to 0.01.
    The peaks around $\omega=\pm 0.27$ correspond to the DWBSs.
    (d) $A_0$ dependence of the low-frequency mode in the domain wall.
    The gray dashed-line $\omega = a A_0 + b$ is calculated from the low-frequency modes by using the least-squares method, where $a=4.26$, $b=6\times 10^{-4}$.
    }
    \label{fig:NG mode}
\end{figure*}
In this appendix, we explicitly show the equation of motion after performing a Fourier transform in the $y$ direction, which is possible thanks to the translational symmetry along the $y$ direction (Fig.~\ref{fig:setup}(a)).
The Fourier transform of the  electron annihilation operator at site $i$ with spin $\sigma$ for the whole Brillouin zone $-\pi < k_y \le \pi$ is given as
\begin{equation}
    c_{i\sigma} = c_{i_x, i_y\sigma} = \dfrac{1}{\sqrt{N_y}} \sum_{k_y} e^{ik_y r_y} c_{i_x, k_y\sigma}.
\end{equation}
The normal and anomalous density matrices introduced in Eq.~\eqref{eq:dGdt} and \eqref{eq:dFdt} are transformed as 
\begin{align}
    \mathscr{G}_{i_x, k_y, \sigma; i_x^\prime, k_y^\prime, \sigma^\prime}
    &=
    \dfrac{1}{N_y} \sum_{r_y, r_y'} e^{ik_y^\prime r_y^\prime -ik_y r_y}
    \mathscr{G}_{i_x, i_y, \sigma; i_x^\prime, i_y^\prime, \sigma^\prime}, \\
    \mathscr{F}_{i_x, k_y, \sigma; i_x^\prime, k_y^\prime, \sigma^\prime}
    &=
    \dfrac{1}{N_y} \sum_{r_y, r_y'} e^{-ik_y^\prime r_y^\prime -ik_y r_y}
    \mathscr{F}_{i_x, i_y, \sigma; i_x^\prime, i_y^\prime, \sigma^\prime}.\hspace{-1em}
\end{align}
We introduce the charge density and SC order parameters diagonalized for $k_y$:
\begin{align}
\begin{split}
    n_{i_x,q,\sigma}
    &= \dfrac{1}{N_y} \sum_{k_y} \Braket{c^\dagger_{i_x,k_y,\sigma} c_{i_x,k_y+q, \sigma}}\\
    &= \dfrac{1}{N_y} \sum_{k_y} \mathscr{G}_{i_x, k_y+q, \sigma : i_x, k_y, \sigma},
\end{split}\\    
\begin{split}
    \Delta^{\mathrm{SC}}_{i_x,q}
    &= \dfrac{1}{N_y} \sum_{k_y} \Braket{c_{i_x,-(k_y+q),\downarrow} c_{i_x,k_y, \uparrow}}\\
    &= \dfrac{1}{N_y} \sum_{k_y} \mathscr{F}_{i_x, k_y, \uparrow : i_x, -(k_y+q) \downarrow}.
\end{split}    
\end{align}
Even in the nonequilibrium dynamics considered in this paper, $q$ takes $0$ or $\pi$ owing to the presence of the CDW order $\bm{Q}=(\pi,\pi)$ and the translational symmetry along the $y$ direction.
As a result, we have $2\times2\times N_x$ mean-fields.
Note again that we set the polarization of the vector potential as $\bm{e}_p=(1,0)$ in order to restrict the photoinduced dynamics to the $x$ direction.
The explicit forms of the equations of motion are 
\onecolumngrid
\begin{equation}
\begin{split}
    &\hspace{-2em}-i\dfrac{d}{dt} \mathscr{G}_{i_x, k_y, \sigma; j_x, k_y', \sigma'}\\
    &=
    J\sum_{\delta = \pm 1}\left[
    (e^{i k_y^{\prime}\delta} - e^{i k_y \delta} )\mathscr{G}_{i_x, k_y, \sigma; j_x, k_y', \sigma'}
    + e^{-iA(t)\delta}\mathscr{G}_{i_x, k_y, \sigma; j_x + \delta, k_y', \sigma'}
    - e^{iA(t)\delta}\mathscr{G}_{i_x + \delta, k_y, \sigma; j_x , k_y', \sigma'}\right]
    \\
%    &-2 J \cos k_y\, \mathscr{G}_{i_x, k_y, \sigma; j_x, k_y', \sigma'} - J(\mathscr{G}_{i_x+1, k_y, \sigma; j_x, k_y', \sigma'} + \mathscr{G}_{i_x-1, k_y, \sigma; j_x, k', \sigma'}) \\
%    &+ 2 J \cos k_y'\, \mathscr{G}_{i_x, k_y, \sigma; j_x, k_y', \sigma'} + J(\mathscr{G}_{i_x, k_y, \sigma; j_x+1, k_y', \sigma'} + \mathscr{G}_{i_x, k_y, \sigma; j_x-1, k_y', \sigma'}) \\
    &\quad+ \sum_{q = 0,\pi}\Bigl\{ U\left[
    (n_{j_x,q,\bar{\sigma'}}
    \mathscr{G}_{i_x, k_y, \sigma; j_x, k_y'+q, \sigma'} - n_{i_x,q,\bar{\sigma'}}
    \mathscr{G}_{i_x, k_y-q, \sigma; j_x, k_y', \sigma'} )
    \right. \\
    & \qquad\qquad\qquad+ \left. ( \delta_{\sigma, \uparrow}  - \delta_{\sigma,\downarrow}
    )  \Delta^{\mathrm{SC}}_{i_x,q}  \mathscr{F}^*_{i_x, -(k_y+q), \bar{\sigma} : j_x, k_y', \sigma'} 
    + ( \delta_{\sigma', \uparrow} - \delta_{\sigma',\downarrow}
    ) \Delta^{\mathrm{SC}*}_{j_x,q} \mathscr{F}_{i_x, k_y, \sigma : j_x, -(k_y'+q), \bar{\sigma^{\prime}}} \right]
    \\
    &\qquad\qquad + V \sum_s \sum_{\ell = j_x,i_x} (2e^{iq}n_{\ell_x,q,s} + n_{\ell_x+1,q,s} + n_{\ell_x-1,q,s}) (\delta_{\ell,j_x} \mathscr{G}_{i_x, k_y, \sigma; j_x, k_y'-q, \sigma'} - \delta_{\ell, i_x}
    \mathscr{G}_{i_x, k_y-q, \sigma; j_x, k_y', \sigma'})\Bigr\}
\end{split}
\end{equation}and
\begin{equation}
\begin{split}
    &\hspace{-2em}-i \dfrac{d}{dt} \mathscr{F}_{i_x, k_y, \sigma; j_x, k_y', \sigma'}\\
    &=
    -J\sum_{\delta=\pm 1}\left[ (e^{ik_y^\prime \delta} + e^{ik_y \delta} - \mu/J) \mathscr{F}_{i_x, k_y, \sigma; j_x, k_y', \sigma'}
    + e^{iA(t)\delta} (\mathscr{F}_{i_x+\delta, k_y, \sigma; j_x, k_y', \sigma'} 
    + \mathscr{F}_{i_x, k_y, \sigma; j_x+\delta, k_y', \sigma'}) \right] \\
    &\quad+\sum_{q=0,\pi} \Bigr\{ U \left[ -n_{i_x,q,\bar{\sigma}}\mathscr{F}_{i_x, k_y-q, \sigma : j_x, k_y', \sigma'}
    - n_{j_x,q,\bar{\sigma'}} \mathscr{F}_{i_x, k_y, \sigma : j_x, k_y'-q, \sigma'} 
    - \delta_{i_x, j_x} \delta_{k_y+k_y',-q}\Delta^{\mathrm{SC}}_{j_x,q} 
    \right.\\&\quad\quad\qquad\qquad\left.
    + \Delta^{\mathrm{SC}}_{j_x,q}(\delta_{\sigma', \downarrow} - \delta_{\sigma', \uparrow}) \mathscr{G}_{i_x, k_y, \sigma : j_x, -(k_y'+q), \bar{\sigma}^\prime}
    + \Delta^{\mathrm{SC}}_{i_x,q}
    (\delta_{\sigma, \uparrow} - \delta_{\sigma, \downarrow}) \mathscr{G}_{j_x, k_y', \sigma' : i_x, -(k_y+q), \bar{\sigma}} \right] \\
    &\qquad\quad\quad
    - V \sum_s \sum_{\ell = j_x,i_x} (2e^{iq}n_{\ell_x,q,s} + n_{\ell_x+1,q,s} + n_{\ell_x-1,q,s}) (\delta_{\ell,j_x} \mathscr{F}_{i_x, k_y, \sigma; j_x, k_y'-q, \sigma'} - \delta_{\ell, i_x}
    \mathscr{F}_{i_x, k_y-q, \sigma; j_x, k_y', \sigma'})\Bigr\}.% \\
%    &\quad + 2\mu \mathscr{F}_{i_x, k_y, \sigma; j_x, k_y', \sigma'}
\end{split}
\end{equation}
\twocolumngrid
Note that $k_y - k_y'=0$, $\pi$ (mod $2\pi$).  

\section{Phase mode bound within the domain wall} \label{sec:phase mode bound in the DW}
In this appendix, we report some results for $\omega_{\mathrm{ext}}=1.5$ which is above the CDW gap $\omega_{\varg}=1.1$.
Under this condition, we find a kind of phase mode within the domain wall. However, the incubation time for the first phase rotation depends on the system size with a single domain wall, and the phase mode does not appear in cases with two or more domain walls. 
These size effects suggest that the phase mode bound in the domain wall would not appear in a large system with multiple domain walls.

Figures~\ref{fig:NG mode}(a) and \ref{fig:NG mode}(b) show the case for a single domain wall.
The real and imaginary parts of the $\Delta_{\mathrm{SC}}$ around the domain wall start to oscillate alternately at around $t\sim 100$ with a finite frequency.
This mode can be regarded as a phase mode bound within the domain wall.
The time-dependence of the quasiparticle population is shown in Fig.~\ref{fig:NG mode}(c). 
The excitation to the conduction band is due to the resonant pumping by the external field. Interestingly, in addition to the transition from the valence continuum to the conduction one, oscillation of the population of the DWBSs at $\omega \simeq \pm 0.27$ is also observed. The frequency of the population dynamics is the same as that of the phase mode in the domain wall. 

We investigated the frequency of the above-mentioned phase mode as a function of $A_0$ (Fig.~\ref{fig:NG mode}(d)).
The way of extracting the frequency of the mode is briefly explained as follows: There are one positive energy and one negative energy bound states for each $k_y$. For each $k_y$, we extract the time dependence of either positive or negative energy of the DWBSs and its population from $t=0$ to $t=500$.
Next, after averaging over each $k_y$, we plot the $A_0$ dependence of the low-frequency mode.
The diamond markers indicate the positions of maximum intensity.
Figure~\ref{fig:NG mode}(d) shows a linear relation between the frequency of the phase mode and the amplitude $A_0$, \red{as in the case of the} Rabi oscillation.

\bibliographystyle{apsrev4-1}
\bibliography{reference}
\end{document}